\documentclass[aps,prb,twocolumn,superscriptaddress]{revtex4-1}

\usepackage{hyperref} 
\usepackage{graphicx}
\usepackage{bm}
\usepackage{amsmath}
\usepackage{amssymb}

\DeclareGraphicsExtensions{.pdf}
\usepackage{xcolor}

\begin{document}

\author{Guangze Chen}
\affiliation{Department of Applied Physics, Aalto University, 02150 Espoo, Finland}

\author{J. L. Lado}
\affiliation{Department of Applied Physics, Aalto University, 02150 Espoo,
Finland}

\title{Impurity-induced resonant spinon zero modes in Dirac quantum spin-liquids}

\begin{abstract}
	Quantum spin-liquids are strongly correlated phases of matter
	displaying a highly entangled ground state. 
	Due to their unconventional nature, finding
	experimental signatures of these states has proven to be a remarkable
	challenge. Here we show that the effects of local impurities can provide
	strong signatures of a Dirac quantum spin-liquid state.
	Focusing on a 
	gapless Dirac quantum spin-liquid state 
	as realized in NaYbO$_2$, we
	show that a single magnetic impurity coupled
	to the quantum spin-liquid state creates a resonant spinon peak at zero
	frequency, coexisting the original Dirac spinons.
	We explore the spatial dependence of this zero-bias resonance, and show
	how
	different zero modes stemming from several impurities interfere.
	We finally address how such spinon zero-mode
	resonances
	can be experimentally probed
	with
	inelastic spectroscopy and
	electrically-driven
	paramagnetic resonance
	with scanning tunnel microscopy.
	Our results put forward
	impurity engineering as a
	means of identifying Dirac quantum spin-liquids with scanning probe
	techniques, 
	highlighting the dramatic impact of magnetic
	impurities in
	a macroscopically 
	entangled many-body ground state.
\end{abstract}

\date{\today}

\maketitle

\section{Introduction}

Quantum spin-liquids\cite{Lee2008,Balents2010,Rau2016} are exotic magnetic
phases of matter, characterized
by strong quantum fluctuations
and frustration,\cite{HwanChun2015} 
lacking magnetic order even at zero temperature.\cite{Anderson1973}
The unique properties of quantum spin-liquids have attracted much research
interest\cite{Broholmeaay0668,RevModPhys.89.025003,Savary_2016},
in particular for their emergent Majorana physics,\cite{Kitaev2006} and their
long-standing relation with unconventional 
superconductivity.\cite{ANDERSON1987,PhysRevX.6.041007}
A variety of compounds showing quantum spin-liquids
physics have been 
identified\cite{Han2012,Fu2015,Powell_2011, PhysRevX.9.031047,RevModPhys.88.041002, takagi2019concept,
PhysRevLett.91.107001, yamashita2009thermal,
PhysRevB.77.104413,PhysRevLett.112.177201, PhysRevLett.98.107204},
including the gapless triangular lattice Dirac quantum spin-liquid in
NaYbO$_2$\cite{PhysRevB.100.144432,Bordelon2019}, 
and different
van der Waals materials.\cite{Law2017,Klanjek2017,Chen2020,Banerjee2016} 
Interestingly, finding gapless Dirac spin-liquids in van der Waals materials would provide a spinon version of graphene Dirac
electrons, opening the door to explore strain gauge fields in 
spinons,\cite{PhysRevB.87.165131,Guinea2009,Levy2010} spinon flat bands
by twist engineering\cite{PhysRevB.82.121407,Bistritzer2011,Cao2018super,Cao2018} and impurity-induced 
spinon resonances.\cite{PhysRevLett.96.036801,GonzalezHerrero2016,PhysRevB.81.233405,doi:10.1080/00018732.2014.927109}

Impurities have been recognized as
a powerful smoking gun to identify exotic
electronic orders.\cite{RevModPhys.81.45} A paradigmatic example of this is the non-magnetic
impurities in unconventional superconductors,\cite{RevModPhys.75.657} where
the
emergence of in-gap states is a well-known
signature of unconventional superconductivity.\cite{PhysRev.131.1553,PhysRevB.60.R749,PhysRevB.62.R11969,PhysRevB.100.014519}
In contrast, conventional s-wave superconductors do not show such in-gap states
in the presence of non-magnetic impurities,\cite{Anderson1959} and only 
magnetic impurities
can give rise to in-gap modes.\cite{RevModPhys.78.373,Shiba1968,Rusinov}
Impurities are also a simple way of imaging the Fermi surface of metals,
by measuring Friedel oscillations with scanning probe techniques.\cite{Sprunger1997,PhysRevB.57.R6858,Weismann2009}
Another paradigmatic example are carbon vacancies\cite{PhysRevLett.92.225502,Yazyev2010,PhysRevB.75.125408,PhysRevB.77.195428,PhysRevLett.93.187202,PhysRevLett.96.036801,PhysRevMaterials.3.084003}
and hydrogen ad-atoms\cite{PhysRevB.77.035427,PhysRevB.96.024403}
in graphene, giving rise to a 
divergent density of states\cite{Yazyev2010,PhysRevLett.96.036801} and 
magnetism.\cite{GonzalezHerrero2016,PhysRevB.75.125408,PhysRevB.96.024403}
In this line, recent experimental advances have demonstrated the
possibility of single atom manipulation in a variety of systems
by means of scanning probe techniques.\cite{Hirjibehedin2007,Otte2008,Loth2010,PhysRevLett.103.107203,Loth2012,Loth2010,Hirjibehedin2006,Liebhaber2019,PhysRevLett.121.196803,Kezilebieke2019,Gross2009,Drost2017,Heinrich2004,Ternes2008,PhysRevLett.119.227206,Kalff2016}
This
motivates whether single atom manipulation
\cite{RevModPhys.91.041001} 
can allow to detect
unique features of quantum spin-liquid states.

\begin{figure}[t!]
\center
\includegraphics[width=\linewidth]{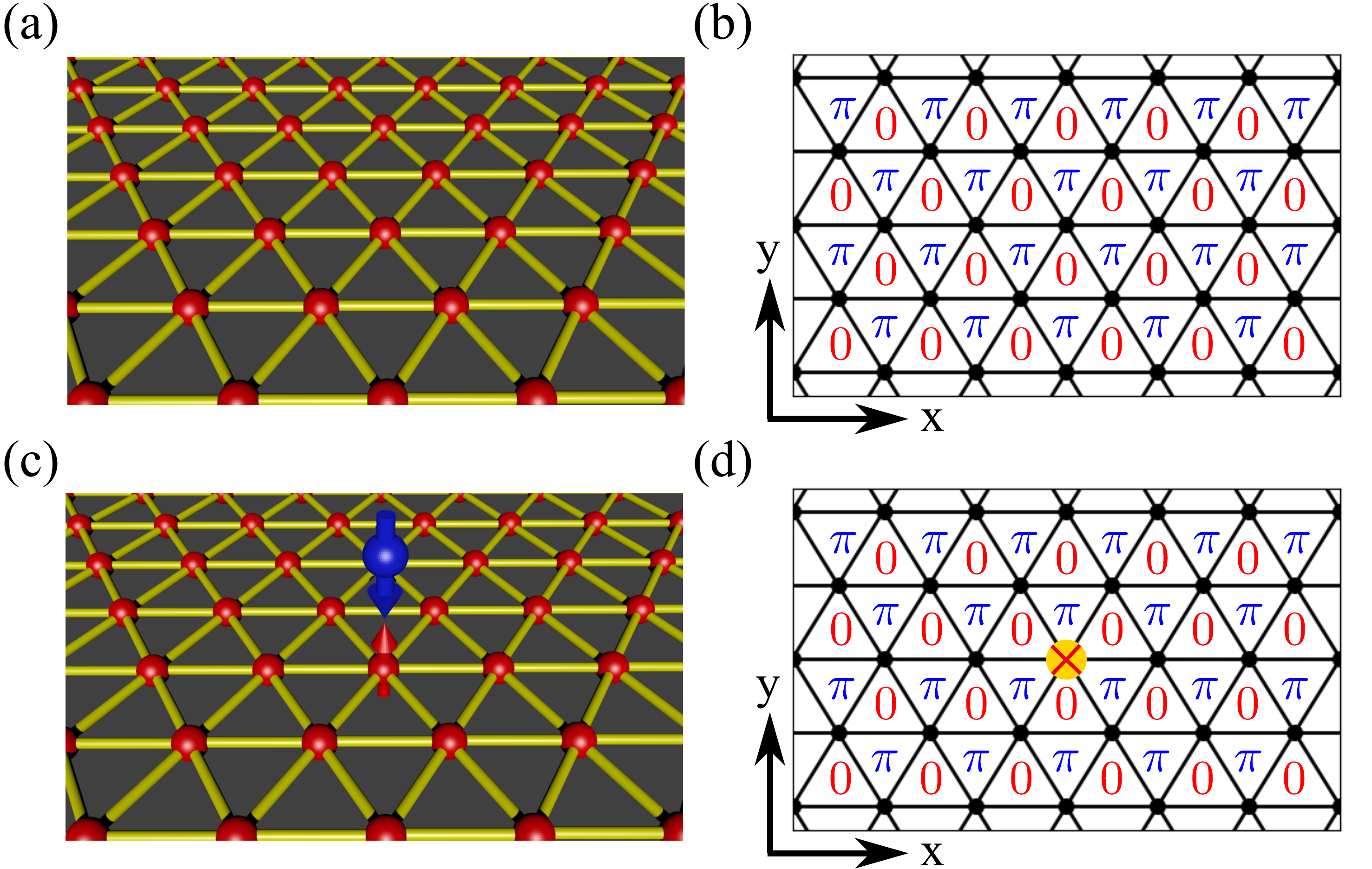}
\caption{(a) Sketch of a Dirac quantum spin-liquid state in the
	triangular lattice, together with (b) the effective
	mean field spinon $\pi$-flux model.
	(c) Sketch of a magnetic moment coupled to the
	triangular Dirac quantum spin-liquid.
	The magnetic impurity (blue)
	couples to the site underneath (red), creating
	a singlet state $|\uparrow \downarrow \rangle - 
	|\downarrow \uparrow \rangle
	$, removing the spin degree of freedom of the coupled
	site from the QSL.
	The combination of the magnetic impurity with the mean-field
	partonic transformation of Eq.\eqref{eq2} give rise
	to an effective model featuring a $\pi$-flux 
	lattice with a vacant site (d).
	}
\label{fig1}
\end{figure}

Here we show that, by depositing individual magnetic atoms on top of a 
Dirac quantum spin-liquid,
spinon resonances can be engineered. We demonstrate that the 
Dirac quantum spin-liquid
ground state develops resonant 
zero modes, and we study the interference
effects between these spinon zero modes. Finally, we show that divergent
spinon density of states can be experimentally probed by means
of inelastic spectroscopy and electrically driven
paramagnetic resonance with scanning tunnel microscope.
Our results put atomically controlled defect engineering 
as a powerful local probe
of Dirac quantum spin-liquid physics, opening up a simple technique
to identify 
fractionalized quantum states of matter with real space measurements.

Our manuscript is organized as follows. In section \ref{sec:single} we show that
single magnetic impurities
create zero energy resonances in a Dirac quantum spin-liquid. In section 
\ref{sec:two} we study the interference effects between different
spinon resonances. In section \ref{sec:detect} we elaborate how such
spinon resonances can be probed by means of scanning tunneling spectroscopy
techniques. Finally, in section \ref{sec:conclusion} we summarize
our conclusions.

\section{Single spin impurity in a Dirac spin-liquid} \label{sec2}
\label{sec:single}

In this section, we show the emergence of resonant zero modes in presence of a
single impurity in a gapless Dirac spin-liquid. We consider two different limiting cases:
(i) a periodic array of impurities with each impurity 
in a unit cell of size
$n\times m$ and (ii) a single impurity in an infinitely
large system.

The spinon excitations in triangular 
spin-liquids [Fig.\ref{fig1}(a)]
such as NaYbO$_2$ are
captured by the $\pi$-flux model on the triangular lattice [Fig.\ref{fig1}(b)]. 
The elementary excitations of the $\pi$-flux state are
Dirac fermions at half-filling. 
A local $S=1/2$ magnetic moment coupled to the 
quantum spin-liquid state
[Fig.\ref{fig1}(c)]
gives rise to a
vacancy in the effective spinon model [Fig.\ref{fig1}(d)]. 
As we will see below, the existence of the magnetic impurity creates
a divergent density of states in the spinon spectra.
For the sake of completeness, we first introduce the spinon properties
of the pristine quantum spin-liquid, and we then move to study the
effect of a magnetic impurity.

\subsection{Spinon excitations in a pristine Dirac spin-liquid}

We start by taking a quantum spin model in a triangular lattice
with the general form
\begin{eqnarray} \label{eq1}
	\mathcal{H} = \sum_{ij}J^{\mu\nu}_{ij} S_i^\mu S^\nu_j
\end{eqnarray}
where $J^{\mu\nu}_{ij}$ are exchange constants
between sites $i,j$ for the spin components
$\mu,\nu$,
and $S^\mu_i$ is the $\mu$ component of the 
spin operator
for the site $i$.
The previous Hamiltonian describes a purely many-body system,
whose exact solution can not be generically found analytically.
The previous model on a triangular lattice
is known to give rise to a quantum spin-liquid state,
when one considers first and second 
neighbor interactions.\cite{PhysRevLett.123.207203,SciPostPhys.4.1.003,PhysRevLett.120.207203}
An approximate solution in a quantum spin-liquid state
can be obtained by performing 
the parton transformation\cite{Savary_2016}
\begin{eqnarray} \label{eq2}
\mathbf{S}=\frac{1}{2}f^{\dag}_\alpha\sigma_{\alpha\beta}f_\beta
\end{eqnarray}
to the model Eq. \eqref{eq1}. The parton transformation separates the 
frozen charge degree of freedom and the free spin degree of
freedom in the quantum spin-liquid state, with $f^{\dag}_\alpha$ and $f_\alpha$ being fermionic
spinon operators with spin-1/2   $\alpha$ satisfying $\sum_\alpha
f^{\dag}_\alpha f_\alpha=1$. At the partonic
mean-field level, and upon the appropriate
regime in the exchange couplings.,\cite{SciPostPhys.4.1.003} the above spin
Hamiltonian gives rise to the $\pi$-flux state\cite{PhysRevB.93.144411,PhysRevLett.123.207203} [Fig.\ref{fig1}(b)]:
\begin{eqnarray} \label{eq3}
H = t\sum_{\langle i,j\rangle}\chi_{ij}f^{\dag}_{i}f_j,
\end{eqnarray}
where $\chi_{ij}=\pm1$ and $t$ are mean-field parameters. 
The $\pi$-flux state hosts
alternating $0,\pi$ fluxes per unit cell. The elementary excitations
of the $\pi$-flux Hamiltonian are spinon
Dirac fermions. 
In the following we show how the presence of a magnetic impurity
modifies the previous picture.

\begin{figure}[t!]
\center
\includegraphics[width=\linewidth]{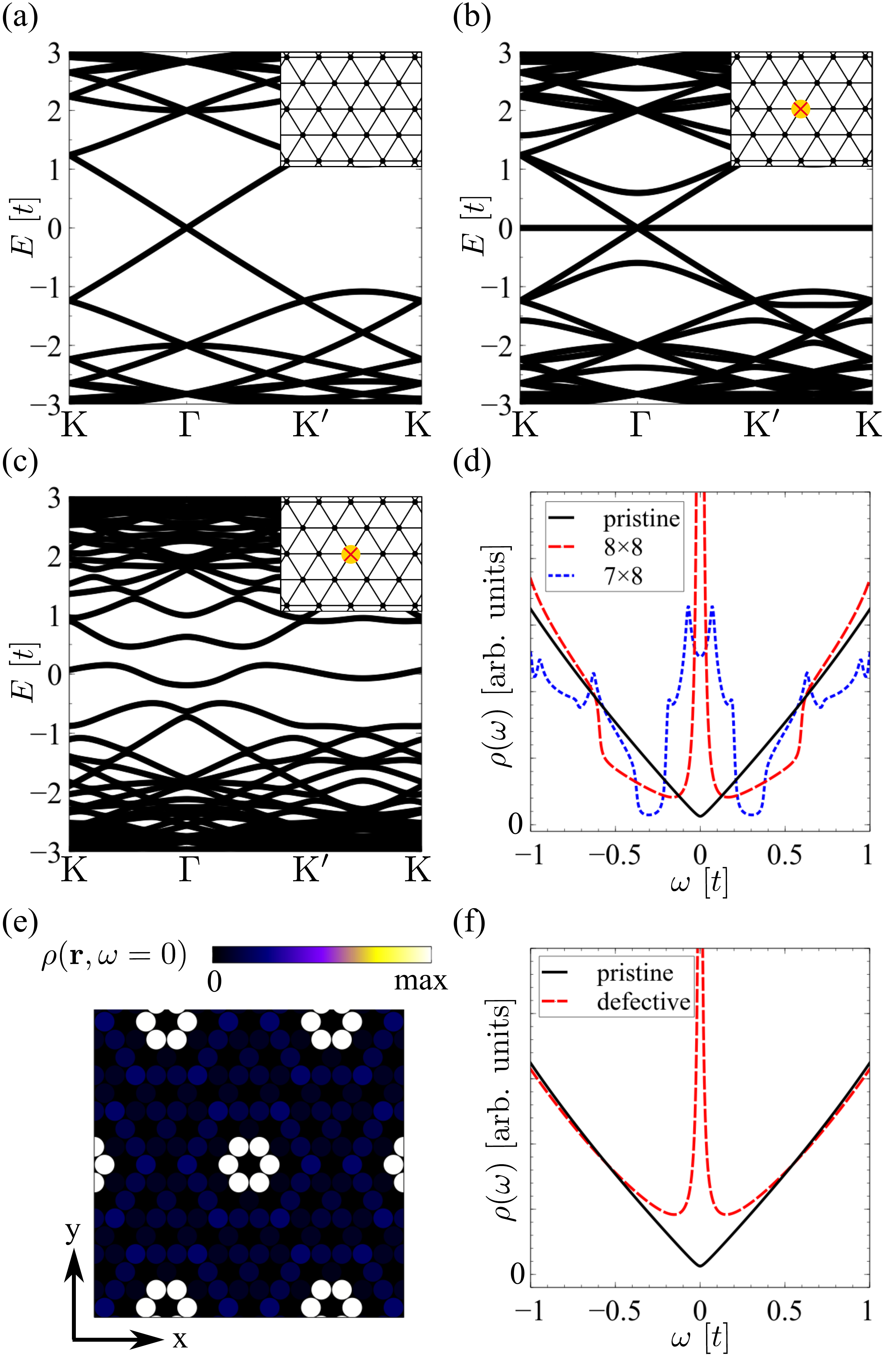}
\caption{Spinon excitations in pristine and defective 
	$\pi$-flux Dirac QSL. Panel (a) shows
	the pristine spinon bandstructure
	of a $\pi$-flux QSL in an $8\times 8$
	unit cell. Panels (b) and (c) show the spinon bandstructure with a periodic array of impurities
	in an $8\times 8$ and a $7\times 8$ unit cell, respectively.
	The insets show the
	configurations of the impurity. Panel (d) shows the 
	DOS corresponding to the three
	cases (a-c). The divergent DOS at zero frequency corresponding to case
	(b) indicates the existence of zero modes, in agreement with the
	bandstructure shown in (b). Panel (e) shows the LDOS at zero frequency
	$\rho(\mathbf r,\omega=0)$ for case (b). The zero modes are localized around the impurities, displaying a pattern with local $C_6$ rotational symmetry. Panel (f) shows the DOS 
	of the pristine QSL, and for a single impurity in an infinite
	QSL.
	}
\label{fig2}
\end{figure}

\subsection{Spinon resonances with periodic impurities}

We now move on to consider the effect of magnetic impurities coupled to
the Dirac quantum spin-liquid state. The total Hamiltonian of the system
is

\begin{equation}
	\mathcal{H} = \mathcal{J}\sum_{k\in \mathcal{K}} 
	\mathbf s_k \cdot \mathbf S_k
	+
	\sum_{ij}J^{\mu\nu}_{ij} S_i^\mu S^\nu_j
\end{equation}
where $\mathbf s_k$ are the spin operators for the different
$S=1/2$ ad-atoms considered,
$\mathcal{K}$ denotes the sites that have an impurity ad-atom on top, and
$\mathcal{J}$ is the antiferromagnetic
exchange coupling between the magnetic ad-atom and the site below.
Taking the limit of strongly coupled magnetic impurity
$\mathcal{J} \gg J^{\mu\nu}_{ij}$, the different sites $k$ will form
a singlet state with the impurity on top, effectively removing the
$S=1/2$ from the quantum spin-liquids compound. As a result, the effective Hamiltonian
in this limit is
\begin{equation}
	\mathcal{H} = 
	\sum_{ij,i\notin \mathcal{K}, j\notin \mathcal{K} }J^{\mu\nu}_{ij} S_i^\mu S^\nu_j,
\end{equation}
an effective triangular model where the sites hosting a magnetic impurity
above disappear from the low energy Hamiltonian. 
Using an analogous spinon replacement as before, we obtain that the
effective model for the spinons becomes
\begin{eqnarray} \label{eq3}
H = t\sum_{\langle i,j\rangle,i\notin \mathcal{K}, j\notin \mathcal{K}}\chi_{ij}f^{\dag}_{i}f_j,
\end{eqnarray}
an effective $\pi-$flux model with impurities determined by the magnetic
ad-atoms deposited.\footnote{We here approximate
that the mean field spinon model does not have non-trivial
reconstructions}
As a result, a magnetic impurity becomes equivalent to a vacancy
in the effective spinon model.
We note that this equivalence holds only for $S=1/2$ impurities,
as higher $S$ impurities would generate a free degree of freedom in each site
even in the limit $\mathcal{J} \gg J^{\mu\nu}_{ij}$.
We also note that given that
the magnetic ad-atoms on top can be moved with a scanning tunnel microscope,\cite{Hirjibehedin2007,Otte2008,Loth2010,PhysRevLett.103.107203,Loth2012,Loth2010,Hirjibehedin2006,Liebhaber2019,PhysRevLett.121.196803,Kezilebieke2019,Gross2009,Drost2017,Heinrich2004,Ternes2008,PhysRevLett.119.227206,Kalff2016}
this would allow to engineer models with an arbitrary number of vacancies
in the effective spinon model.

We now explore the spectra of this defective quantum spin-liquid state.
When considering a periodic array of impurities in unit cells of
size $n\times
m$, the Bloch Hamiltonian can be
used to compute the bandstructure and the 
density of states (DOS). Compared to the bandstructure of
the pristine $\pi$-flux state, a flat band at zero energy arises for $n$ even
[Fig.\ref{fig2}(b)], and a wiggly band near zero energy arises for $n$ odd
[Fig.\ref{fig2}(c)]. In both cases, the DOS at zero frequency shows a dramatic
increase
[Fig.\ref{fig2}(d)]. 
The dispersive zero mode for odd $n$ stems from the self-interaction
effects of the zero mode, which are absence in the $n$ even case.
As it is expected, as $n$ is increased, the zero mode band
becomes flatter even for odd $n$ due to the decrease self-interaction
between replicas.
For a finite unit cell,
the DOS diverges at zero frequency
only when $n$ is even, indicating the existence 
of resonant zero modes. The nature
of the zero mode can be analyzed by looking
at the 
local density of states (LDOS) defined as
$\rho (\mathbf r,\omega) = \Im 
\left (\langle \mathbf r | [\omega - H - i0^+]^{-1} | \mathbf r \rangle \right )$.
In particular, the LDOS at zero
frequency $\rho(\mathbf r,\omega=0)$ shows that the zero 
modes are localized around the
impurities [Fig.\ref{fig2}(e)], showing
a pattern with local $C_6$ rotational symmetry. 
Interestingly, the zero modes are 
mainly localized through sites that are 
odd number of bonds straight away from the impurity.
The previous calculation relied on assuming a periodic pattern of
impurities. Experimentally, the simplest scenario will be depositing 
a single impurity in an infinite quantum spin-liquid. In the
following we will deal with this idealized case, showing that the results
are qualitatively similar to the periodic impurity pattern considered above.

\subsection{Spinon resonances for a single impurity in an infinite Dirac
spin-liquid}
We now move on to consider a single impurity coupled to the quantum spin-liquid.
In the case of a single impurity in an infinite system, translational symmetry
is broken and a Bloch Hamiltonian can not be defined. To deal with 
this inhomogeneous infinite problem, we compute exactly the spectral function
close to the impurity using a Green's function
embedding
method.\cite{PhysRevB.82.085423,lado2016unconventional,PhysRevB.96.024403}
For the sake of completeness, we now summarize the essence of the method.
For a unit cell containing
the impurity, the Green's function in this unit cell can be 
written using Dyson's equation as:
\begin{eqnarray} \label{eq4}
G(\omega)=(\omega-H'-\Sigma(\omega))^{-1},
\end{eqnarray}
where $H'$ is the Hamiltonian of the unit cell and $\Sigma(\omega)$ is the
self-energy due to the coupling of the unit cell to the rest of the infinite
pristine system. The
impurity does not influence $\Sigma(\omega)$ since it does not change the
hoppings that couple the unit cell to the rest of the system. Therefore, in
the absence of the impurity, the Green's function of the pristine
unit cell coupled to the infinite system is:
\begin{eqnarray} \label{eq5}
G_0(\omega)=(\omega-H_0-\Sigma(\omega))^{-1},
\end{eqnarray}
where $H_0$ is the Hamiltonian of the pristine unit cell. Since the whole
system is now pristine, this Green's function can also be computed by:
\begin{eqnarray}
	G_0(\omega)=\frac{1}{(2\pi)^2}\int d^2 \mathbf k (\omega-H_{\mathbf k} -i 0^+)^{-1},
\end{eqnarray}
where $H_{\mathbf k}$ is the Bloch Hamiltonian 
associated to Eq.\eqref{eq3} on this unit cell. Using
Eq.\eqref{eq5}, the self-energy can be computed as
$\Sigma(\omega)=\omega-H_0-G_0^{-1}(\omega)$, and the Green's function of the defective
unit cell can be solved with Eq.\eqref{eq4}. The DOS is thus
$\rho(\omega)=-\frac{1}{\pi}\Im G(\omega)$, which diverges at zero frequency
for a defective unit cell [Fig.\ref{fig2}(f)], indicating the existence of
zero modes. As a result, a single impurity in the quantum spin-liquid
state also gives rise to a zero mode, as anticipated
from the calculations in periodic arrays of Figs.\ref{fig2}(b,c,d,e).
We note that the embedding method presented above would allow
to compute an arbitrary cluster of impurities
coupled to the quantum spin-liquid, as we will address in the next section.

We have thus verified the existence of zero modes when there is
(i) a periodic array of impurities with one in an $n\times m$ unit cell with
$n$ even and (ii) a single impurity in an infinite system. 
This resonant zero mode stems from the vacancy
boundary conditions in a Dirac system, which is known
to give rise to zero modes in other Dirac systems.\cite{PhysRevLett.96.036801}
We finally note that so far we have focused on single impurities, yet
when several impurities are put together, the different zero modes are
expected to give rise to interference effect. We address this interference
phenomenon in the next section.

\section{Interaction between zero modes}
\label{sec:two}

In this section, we investigate the interaction between zero modes by
considering the case when more impurities are present. For the sake
of simplicity, we first consider interference between two and three
impurity sites, and observe a large dependence of the net number of zero modes on their
relative location.
We then generalize our
discussion to the thermodynamic limit when a certain
density of impurities is randomly distributed,
giving rise
to a sublinear increase of DOS at 
zero frequency with respect to the impurity density due to interference effects.

\subsection{Spinon zero mode interference between
individual resonances}\label{sec3}

\begin{figure}[t!]
\center
\includegraphics[width=\linewidth]{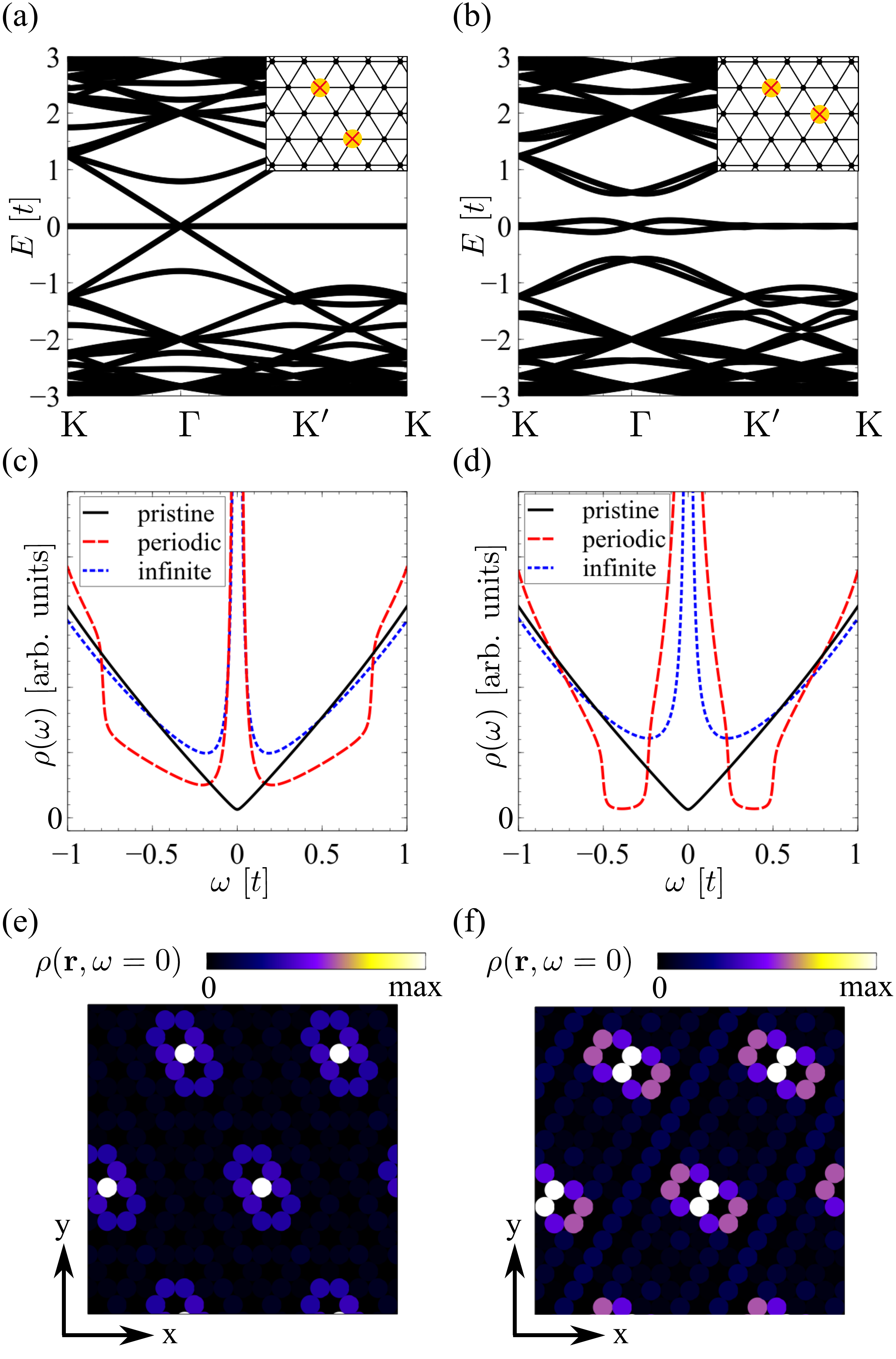}
\caption{Resonant zero modes with two impurities in the
	$\pi$-flux state. Panels (a) and (b) show the
	bandstructure of the $\pi$-flux  QSL state with a periodic array of two
	impurities that are (a) two
	bonds away from each other (b) next-nearest to each other in unit cells
	of size $8\times 8$. The insets show the configurations of the
	impurities. Panels (c) and (d) show the DOS
	of the $\pi$-flux QSL state with periodic impurities and impurities in an infinite system. The 
	impurity configurations for (c) and (d) are shown in (a) and
	(b), respectively. Panels (e) and (f) show 
	the LDOS at zero frequency $\rho(\mathbf
	r,\omega=0)$ of the
	$\pi$-flux state for cases (a) and (b), respectively.}
\label{fig3}
\end{figure}

We first consider the case of two impurities in the
Dirac quantum spin-liquid, for both a periodic array and a single cluster of
impurities in an otherwise pristine system Dirac QSL. 
In the periodic case, we consider two impurities per unit cell of size $n\times m$ with $n$ even in order to observe interference between zero modes, as with $n$ odd there is no zero mode in the single-vacancy case (see section \ref{sec:single}).
We find the relative position between the impurities strongly
impacts the overall zero modes.
We observe that
only when the two impurities are
an even number of bonds straight
away from each other, will there be two zero modes. In
particular, 
we show in 
Fig.\ref{fig3}
the bandstructure [Figs.\ref{fig3}(a,b)], 
DOS [Figs.\ref{fig3}(c,d)], and 
LDOS [Figs.\ref{fig3}(e,f)] of zero modes for the cases (i)
when the two impurities are two bonds away from 
each other [Figs.\ref{fig3}(a,c,e)] and (ii) when the two
impurities are next-nearest to each other 
[Figs.\ref{fig3}(b,d,f)].
For case
(i), a flat band arises at zero energy in the presence of the two impurities
[Fig.\ref{fig3}(a)], whereas in case (ii) wiggly bands at zero 
energy arise
instead [Fig.\ref{fig3}(b)]. 
It is also interesting to note that, in the periodic case, for case
(i) the dispersive bulk states remain gapless [Fig.\ref{fig3}(a)], whereas
for case (ii) they are gapped out [Fig.\ref{fig3}(b)]. 
In both cases, the 
DOS at zero frequency diverges
for periodic impurities and impurities in 
an infinite system [Figs.\ref{fig3}(c,d)].
The LDOS at zero frequency for both cases are 
similar to the summation of LDOS of
zero modes for the two impurities [Figs.\ref{fig3}(e,f)].

The previous picture is dramatically different if the two impurities
were put just next to each other, in which case no zero modes appear
in the system. In this situation, the impurity states created by each impurity
give rise to a bonding-antibonding splitting, lifting both
the spinon resonance from zero frequency.
The dependence of the existence of zero modes on the relative position between
impurities can be understood by starting with the spatial of the zero mode resonant state
 [Fig.\ref{fig2}(e)]. 
Let us now think in a perturbative way, 
in which a second vacancy can be understood as the limit
where a local onsite energy is ramped up to $\pm \infty$.
When an additional 
impurity is added to a site
where the impurity zero mode is finite,
perturbation theory suggests that the zero mode
will be lifted from zero energy.
In contrast,
when the second impurity is 
added to a site where the LDOS
at
zero frequency vanishes, 
perturbation theory would suggest that the original mode remains at zero.
We note that the previous picture is just perturbative and does not 
quantitatively
account for the true double impurity nor the number of zero modes, 
but it provides a simple argument 
to rationalize the persistence of zero modes.

\begin{figure}[t!]
\center
\includegraphics[width=\linewidth]{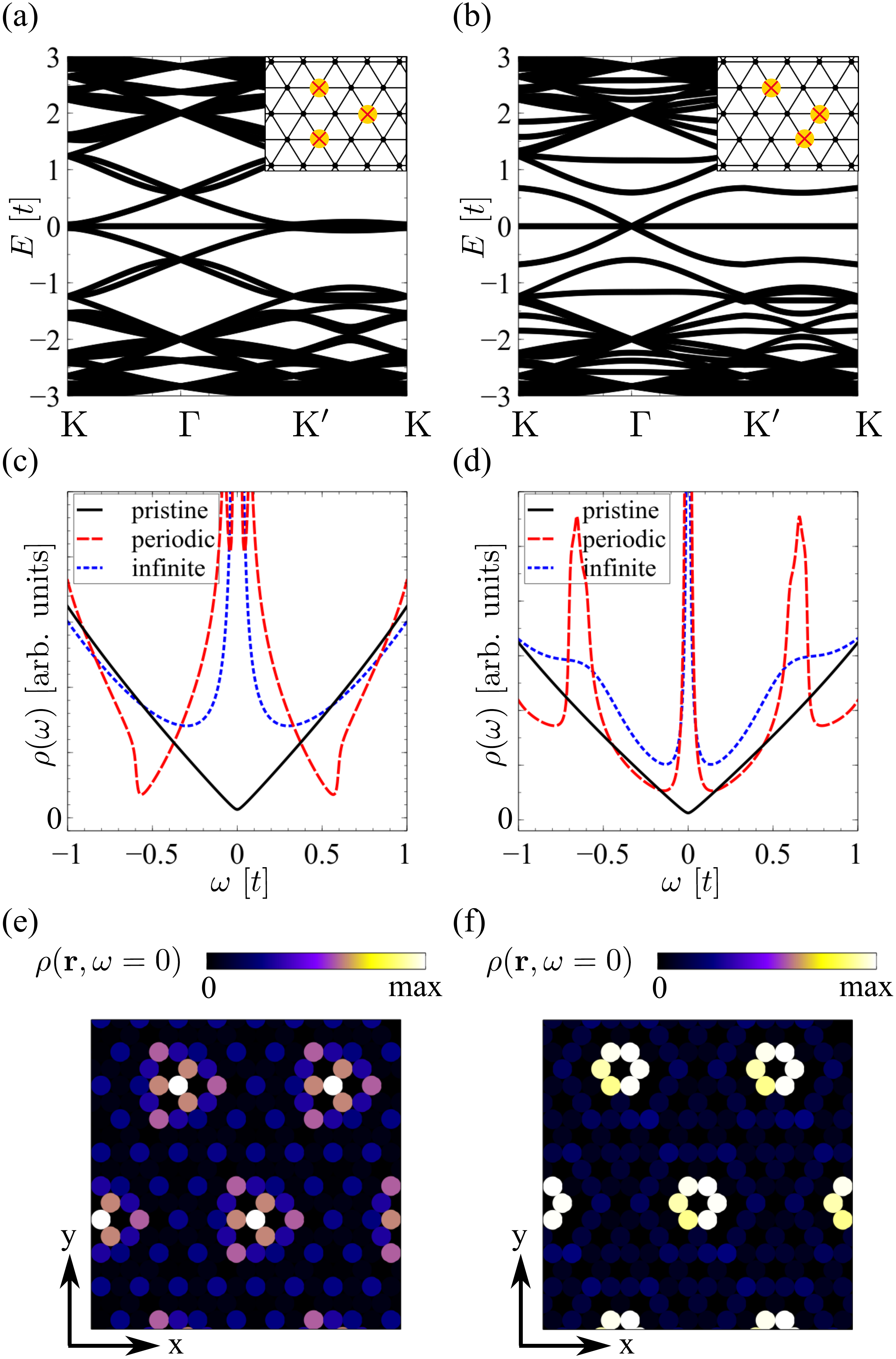}
\caption{Resonant zero modes with three impurities in the
	$\pi$-flux state. Panels (a) and (b) show
	the bandstructure of the $\pi$-flux QSL state with three impurities of
	different configurations shown in the insets in a unit cell of size
	$8\times 8$. Panels (c) and (d) show the DOS
	of the $\pi$-flux QSL state with periodic impurities and impurities in an infinite system. The 
	impurity configurations for (c) and (d) are shown in (a) and
	(b), respectively. Panels (e) and (f) show the LDOS at zero
	frequency $\rho(\mathbf r,\omega=0)$ of the
	$\pi$-flux state for cases (a) and (b), respectively.}
\label{fig4}
\end{figure}

We now move on to consider the case of three impurities
as shown in Fig. \ref{fig4}. 
We will focus on arrangements that still give rise to zero modes.
We proceed
in an analogous way, by showing the bandstructure [Figs. \ref{fig4}(a,b)],
DOS [Figs. \ref{fig4}(c,d)] and LDOS [Figs. \ref{fig4}(e,f)].
We first focus on the case in which impurities are arranged in a $C_3$
symmetric fashion [Figs. \ref{fig4}(a,c,e)], in a fashion similar to 
the double impurity of [Figs. \ref{fig3}(a,c,e)]. As is shown in 
Fig. \ref{fig4}(c), a zero mode appears even with three impurities
close to each other, giving rise to a zero mode
with $C_3$ rotational symmetry. 
The interference of the three impurities
is again highly sensitive to their relative position.
In particular, by taking the limiting case of two
impurities next to each other and one further apart
[Figs. \ref{fig4}(b,d,f)], we observe a zero frequency peak
surrounded by two peaks at positive and negative frequency
[Fig. \ref{fig4}(d)].
The two peaks above and below zero can be understood as the bonding
and antibonding impurity resonances associated to the closest impurities,
whereas the remaining zero frequency peak will be be associated to the
remaining impurity.
This is also shown in Fig. \ref{fig4}(f), where it can be seen
that the zero mode is located around the remaining impurity, with
a small $C_6$ symmetry breaking induced by the other two impurities.
These results highlight that the interaction between zero modes
created by different impurities will give rise to non-trivial
interference effects. In particular, this will give rise to a density
of zero modes sublinear with the impurity density as we address below.

\subsection{Thermodynamic limit and zero mode quenching}

\begin{figure}[t!]
\center
\includegraphics[width=\linewidth]{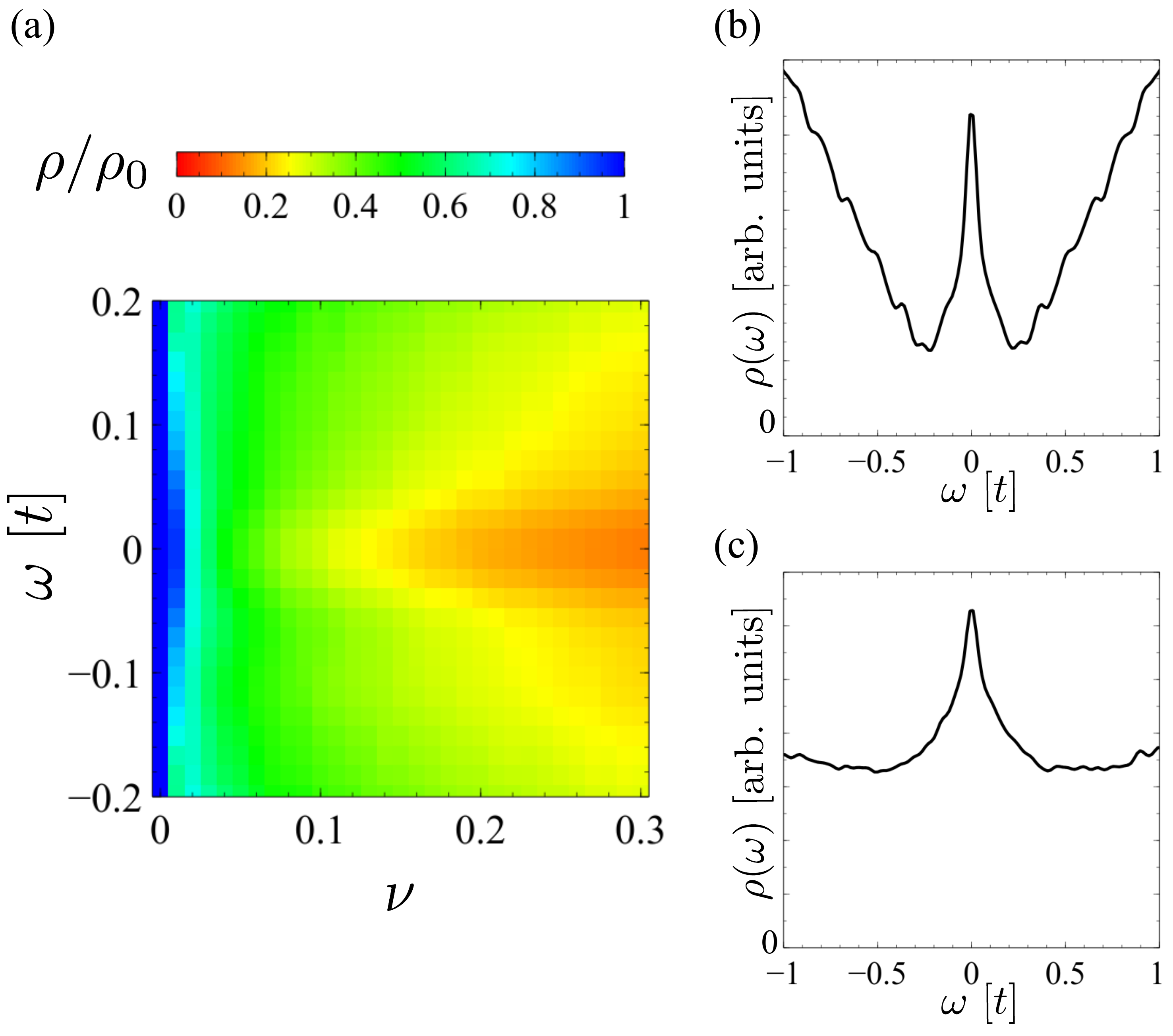}
\caption{(a) Ratio of the computed DOS over the expected DOS
        in the absence of interference effects
	$\rho(\nu,\omega)/\rho_0(\nu,\omega)$ for different impurity density
	$\nu$
	in the $\pi$-flux state. The sublinear increase of such ratio with respect
	to $\nu$ at $\omega=0$ indicates the existence of interference effects
	between different zero modes.
	Panel (b) shows the DOS for the $\pi$-flux state at impurity
	density $\nu=0.01$, a low
	impurity concentration showing negligible interference.
	Panel (c) shows the DOS for the $\pi$-flux state at impurity
	density $\nu=0.25$,
	showing a broader peak stemming from the interference and quenching between vacancy states.}
\label{fig5}
\end{figure}

We now address the emergence of zero modes in a disordered system
with a varying number of impurities.
In particular, we will show how the interference
of zero modes can be observed by tracking the spinon
DOS near zero frequency for
different impurity density. 
We start by discussing an idealized case in which there
are no interference effect.
If there were no interactions between different
zero modes, the density of zero
modes should increase linearly with respect 
to the impurity density,
assuming that each new impurity would create a new zero mode.
In this idealized case,
the expected
density of states 
$\rho_0(\nu,\omega)$
at a certain impurity density $\nu$
would fulfill
\begin{eqnarray}
	\lim_{\omega\rightarrow 0}\rho_0(\nu,\omega)= \lim_{\omega\rightarrow
	0} 
	\left (
	\rho(0,\omega)+\frac{\nu}{\nu_0}\rho(\nu=\nu_0,\omega)
	\right )
\end{eqnarray}
with $\omega$ near 0, $\nu_0$ a small finite impurity concentration.
and $\rho(\nu,\omega)$ the true density of states of the system
computed exactly. In the following we take $\nu_0=0.01$, and we verify
that our results remain qualitatively similar with other small values.
We now compute 
the density of
states $\rho(\nu,\omega)$ for different impurity densities and at
different energy 
using kernel 
polynomial method (KPM)\cite{RevModPhys.78.275}, 
and averaging over impurity distributions. We consider 
random impurity distribution with density $\nu$
from $0.01$ to $0.3$, where $\nu=1$
would mean vacancies in every site.
We take a unit cell of size $50\times 50$, and show
$\rho(\nu,\omega)/
\rho_0(\nu,\omega)$, the ratio
of the computed exact DOS 
over the expected DOS in the absence of interference, for different 
$\nu$ and $\omega$ [Fig.\ref{fig5}]. 

In the ideal case in which impurity interferences are negligible,
the ratio 
$\rho(\nu,\omega)/
\rho_0(\nu,\omega)$
would be one for $\omega\rightarrow 0$,
as it is observed at small $\nu$ in Fig.\ref{fig5}(a).
This is easily rationalized by taking into account that
at small concentrations, interference effects between impurities
are statistically unlikely, and therefore the system behaves
as if each impurity is isolated. 
This is also seen by inspecting the disorder
average DOS for $\nu=0.01$ as shown
in Fig.\ref{fig5}(b), which resembles the result obtained
for a single impurity in an infinite system
shown in Fig.\ref{fig2}(d).
This
situation dramatically changes as the impurity concentration
increases, as can be observed for large values of $\nu$
in Fig.\ref{fig5}(a).
In particular, it is shown in Fig.\ref{fig5}(c) the DOS for 
$\nu=0.25$, highlighting that the zero mode peak has become less 
sharp in comparison with the bulk states.

It is finally worth to note that for larger impurity
concretations, the spinon 
ground state may suffer a reconstruction
in a real experiment
and lose its Dirac nature, and therefore impurity
interference effects are better experimentally
explored at low impurity densities.
This brings up the question on how such zero energy resonances
in the spinon spectra can be experimentally detected,
which is addressed in the next section.

\section{Experimental detection of spinon zero modes}
\label{sec:detect}

\begin{figure}[t!]
\center
\includegraphics[width=\linewidth]{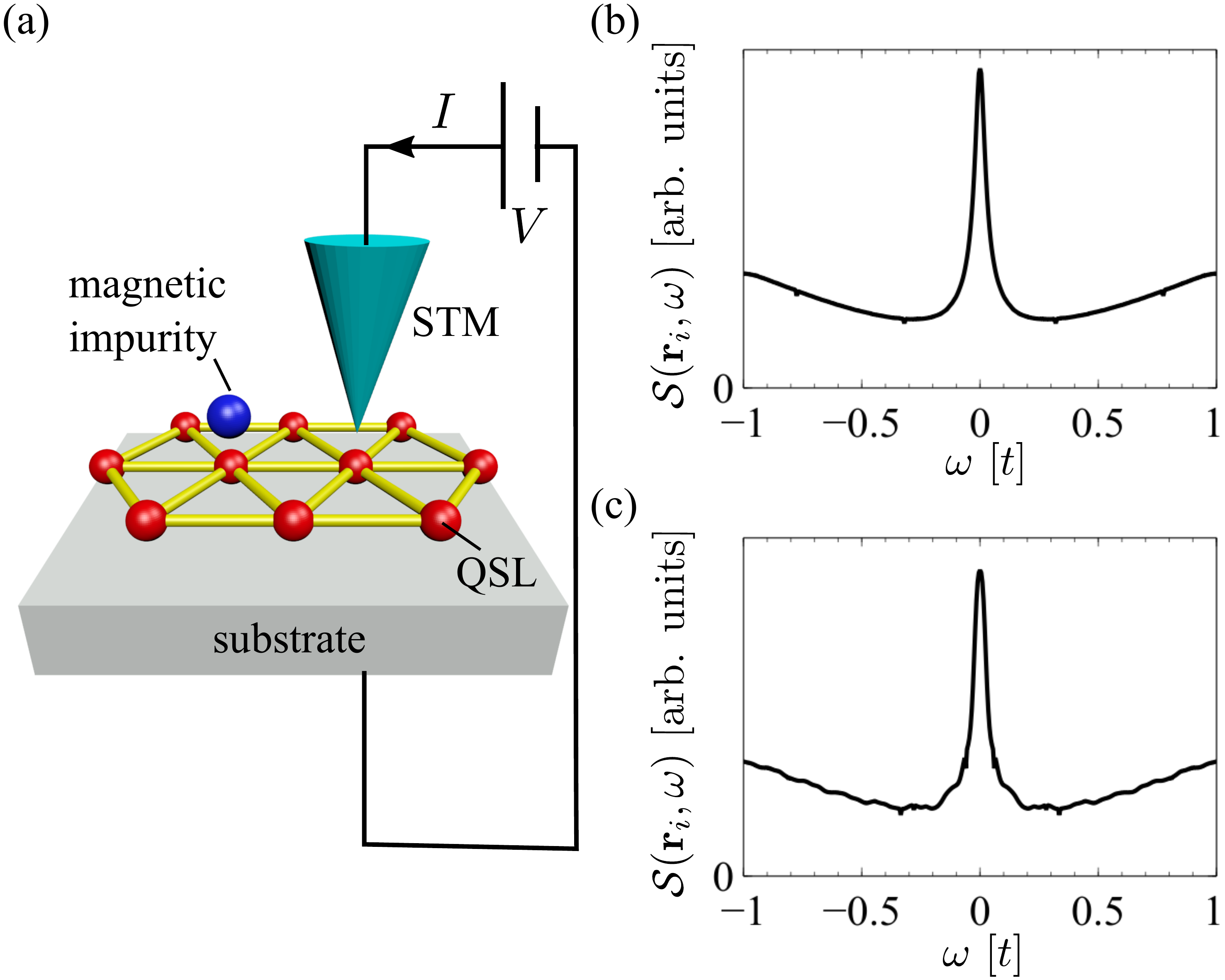}
\caption{(a) Sketch of the experimental setup to measure
the resonant Dirac spinons close to the impurity by means of 
inelastic spectroscopy or electrically driven paramagnetic resonance.
	Panels (b) and (c) show the 
	local spin structure factor $\mathcal{S} (\mathbf{r}_i,\omega)$,
	computed on the site near the impurity, for (b) a single impurity in an infinite
	system and for (c) a single impurity in a large
	finite system (with $100\times 100$ sites).
	It is observed that a zero bias peak appears, which is associated
	with the original divergent spinon density of states at zero
	frequency.
	}
\label{fig6}
\end{figure}

We now consider the potential experimental signatures of these spinon
zero modes. 
In particular, scanning tunnel spectroscopic techniques have been demonstrated
to be very well suited to detect quantum 
spin excitations,\cite{RevModPhys.91.041001} 
as demonstrated in a variety of experiments showing atomic-scale
magnons,\cite{Spinelli2014} quantum critical transitions,\cite{Toskovic2016} and quantum transitions
in nanomagnets.\cite{PhysRevLett.119.227206}
In particular, two different techniques can be used to probe
magnetic excitations with STM: inelastic spectroscopy\cite{Heinrich2004,PhysRevLett.102.256802,Loth2010}
and electrically driven paramagnetic resonance.\cite{Natterer2017,Bae2018,Willke2018adv,Willke2019,Willke2019NN,PhysRevResearch.2.013032,Baumann2015,Willke2018,PhysRevB.96.205420,Willke2018,PhysRevResearch.1.033185}
Inelastic resonance experiments rely on measuring current
versus voltage between the tip and the sample, and identifying steps in the
differential conductance $dI/dV$. These steps are associated
with inelastic processes in which an electron tunnels creating an
excitation, namely a phonon\cite{PhysRevB.69.121414} or spin excitation.\cite{Spinelli2014} 
In particular, neglecting phonon contributions at small biases,
inelastic steps will appear as peaks in the 
${\text{d}^2I}/{\text{d}V^2}$ and are proportional to the spectral
function of spin excitations\cite{Spinelli2014}

\begin{equation}
	\label{eq:dyn}
	{\text{d}^2I}/{\text{d}V^2} \sim 
	\langle GS |S^+_i \delta (\omega - \mathcal{H}+E_{GS}) S^-_i | GS \rangle
\end{equation}
where $|GS\rangle$ is the many-body ground state and $E_{GS}$ is the
many-body ground state energy. The quantity in Eq. \ref{eq:dyn} is 
proportional to the so-called
spin structure factor $\mathcal{S}(\mathbf{r}_i,\omega)$, which can
be understood as the magnon density of states in a ferromagnet, or the
$\Delta S =1$ excitations in a generic spin system. 
In the particular
case of a quantum spin-liquid state, $\Delta S =1$ involves creating
two-spinon excitations, and as
a result provides information about the two-spinon spectral function.
Furthermore, besides
inelastic spectroscopy, the spin structure
factor can be accessed by electrically driven
paramagnetic resonance with STM.\cite{Baumann2015}
This technique has further
demonstrated to allow for measuring spin excitations with a resolution
not limited by temperature,\cite{Natterer2017,Baumann2015,Willke2018}
turning it into a well suited technique to probe
the low energy scales expected in quantum spin-liquid systems.

In the partonic spinon language, the spin structure factor will be
proportional to the density-density response function of the spinons.
We will compare our results between 
a single
impurity coupled to an infinite and otherwise pristine QSL
computed with the embedding method [Fig.\ref{fig6}(b)],
and a single impurity in a finite
large system computed with the KPM\cite{RevModPhys.78.275} [Fig.\ref{fig6}(c)]. 
It is important to note that, in the following
and for the
sake of simplicity, we will be neglecting
gauge fluctuations. Within this approximation, the spin 
structure factor becomes
$
	\mathcal{S}(\mathbf{r}_i,\omega) \sim 
	\sum_{n,n'}\frac{f_{n}-f_{n'}}{\omega+\varepsilon_{n}-\varepsilon_{n'}+i\eta}
	\psi^*_{n}(\mathbf{r}_i)\psi_{n'}(\mathbf{r}_i)\psi_{n}(\mathbf{r}_i)\psi^*_{n'}(\mathbf{r}_i),
	$
where
$\psi_{n}$ is the $n$-th spinon eigenstates with energy $\varepsilon_{n}$,
and $f_{n}$ is the Fermi-Dirac distribution.
It is now convenient to rewrite $\mathcal{S}(\mathbf{r}_i,\omega)$
in terms of the local spectral function
$
\rho(\mathbf{r}_i,\mathbf{r}_i,\omega)=\sum_n\psi^*_{n}(\mathbf{r}_i)\psi_{n}(\mathbf{r}_i)\delta(\omega-E_n),
$
so that the spin structure factor becomes
\begin{multline} \label{eq10}
	\mathcal{S}(\mathbf{r}_i,\omega) \sim 
	\int\text{d}\omega_1\text{d}\omega_2\frac{\rho(\mathbf{r}_i,\mathbf{r}_i,\omega_1)\rho(\mathbf{r}_i,\mathbf{r}_i,\omega_2)}{\omega+\omega_1-\omega_2+i\eta}
\times
	\\
	\left(f(\omega_1)-f(\omega_2)\right).
\end{multline}
The local spin structure factor is computed for the site near the impurity
[Fig.\ref{fig6}(b)], where the Green's function is computed using the
embedding method introduced in section \ref{sec2}. Alternatively, we also show
the
spin structure factor computed for a finite quantum spin-liquid system
with 10000 sites [Fig.\ref{fig6}(c)] using the KPM\cite{RevModPhys.78.275}. In both instances, it is observed a zero bias
peak, which is associated to the divergent density of states of 
the spinon excitations.

The previous result highlights that the spinon zero mode
resonances will appear as a divergent peak at small biases.
In a similar fashion, the different arrangements between magnetic impurities
will give rise to spectras resembling a self-convolution of the spinon density
of states. This feature may allow to distinguish Dirac spin-liquid states
from another generic kind of magnetically ordered state, as
resonant-like zero bias peak for $S=1/2$ will not appear for a generically
coupled magnetic state. We finally note that the previous picture relies
on assuming that the tunneling signal stems solely
from spin flip processes, and neglects orbital or Kondo-like
transitions that can be present in the real setup.

\section{Conclusion}\label{sec_7}
\label{sec:conclusion}
We have shown that individual magnetic $S=1/2$ impurities
coupled to a Dirac quantum spin-liquid state,
as realized in NaYbO$_2$, give rise
to a divergent spinon density of states at zero frequency.
The emergence of such zero modes is associated with the
low energy
Dirac nature of the spinon excitations, and as a result provides
a simple spectroscopic signature distinguishing
Dirac spin-liquids from generic gapless
Dirac liquids with a finite
Fermi surface.
We have shown that such spinon zero modes give rise
to a zero frequency divergence
in the spin structure factor,
that can be measured by means of
inelastic spectroscopy
and electrically-driven paramagnetic resonance 
with scanning tunnel microscopy.
Interestingly, although the emergence of 
zero-bias peaks in inelastic spectroscopy due to
a magnetic impurity is commonly associated with Kondo
physics, the phenomena presented relies on single
particle spinon physics, and it is therefore not related
with a spinon-induced Kondo state.
Our results put forward impurity engineering by scanning probe techniques
as a simple method to probe quantum spin-liquid
physics by a local real space measurement.

\begin{acknowledgments}
	We acknowledge the computational resources provided by the Aalto
	Science-IT project.
	We thank P. Liljeroth for useful discussions.

\end{acknowledgments}

\bibliographystyle{apsrev4-1-etal-title}
\bibliography{biblio}
\bibliographystyle{apsrev4-1}

\end{document}